\documentclass[article]{llncs}
\usepackage{graphicx}
\usepackage{amsmath}
\usepackage{amssymb}
\usepackage{longtable}
\usepackage{hyperref}
\usepackage{array}
\usepackage[T1]{fontenc} 

\usepackage[a4paper, total={6in, 8in}]{geometry}

\usepackage[htt]{hyphenat}

\usepackage[square,sort,comma,numbers]{natbib}
\DeclareMathOperator*{\argmin}{argmin}

\begin{document}

\title{Simulation-based Inference for Exoplanet Atmospheric Retrieval: Insights from winning the Ariel Data Challenge 2023 using Normalizing Flows}

\author{Mayeul Aubin\inst{1,2}$^\ast$%\orcidID{0009-0004-4673-3864}
\and Carolina Cuesta-Lazaro\inst{3,4}
\and Ethan Tregidga\inst{1,5}
\and Javier Via\~{n}a\inst{6}
\and Cecilia Garraffo\inst{1,7} 
\and Iouli E. Gordon\inst{8} 
\and Mercedes L\'opez-Morales\inst{8}
\and Robert J. Hargreaves\inst{8}
\and Vladimir Yu. Makhnev\inst{8}
\and Jeremy J. Drake\inst{7}
\and Douglas P. Finkbeiner\inst{3}
\and Phillip Cargile\inst{9}}
%\institute{Center for Astrophysics $|$ Harvard ${\rm \&}$ Smithsonian, 60 Garden Street, Cambridge, MA 02138, USA
%1 -- AstroAI
\institute{AstroAI, Center for Astrophysics $|$ Harvard ${\rm \&}$ Smithsonian, 60 Garden Street, Cambridge, MA 02138, USA
%2 -- Ecole Polytechnique
\and Ecole Polytechnique, Rte de Saclay, 91120 Palaiseau, France \\
$^\ast$ Corresponding author: \email{m.mayeul.aubin@gmail.com}
%3 -- ITC
\and Institute of Theory and Computation, Center for Astrophysics $|$ Harvard ${\rm \&}$ Smithsonian, 60 Garden Street, Cambridge, MA 02138, USA
%4 -- Carolina MIT
\and Department of Physics, Massachusetts Institute of Technology, Cambridge, MA 02139, USA
%5 -- Southampton
\and University of Southampton, University Road, Southampton, SO17 1BJ, UK
%6 -- Javier MIT
\and Kavli Institute for Astrophysics and Space Research $|$ Massachusetts Institute of Technology. 77 Massachusetts Avenue, Cambridge, MA 02139, USA
%7 -- HEA
\and  Division of High Energy Astrophysics, Center for Astrophysics $|$ Harvard ${\rm \&}$ Smithsonian, 60 Garden Street, Cambridge, MA 02138, USA
%8 -- AMP
\and Division of Atomic and Molecular Physics, Center for Astrophysics $|$ Harvard ${\rm \&}$ Smithsonian,  60 Garden Street, Cambridge, MA 02138, USA
%9 -- OIA
\and Division of Optical and Infrared Astronomy, Center for Astrophysics $|$ Harvard ${\rm \&}$ Smithsonian, 60 Garden Street, Cambridge, MA 02138, USA 
}

\date{July 2023}

\maketitle

\begin{abstract}

Advancements in space telescopes have opened new avenues for gathering vast amounts of data on exoplanet atmosphere spectra. However, accurately extracting chemical and physical properties from these spectra poses significant challenges due to the non-linear nature of the underlying physics. 

This paper presents novel machine learning models developed by the AstroAI\footnote{AstroAI is hosted by the Center for Astrophysics $|$ Harvard ${\rm \&}$ Smithsonian}\footnote{\url{https://astroai.cfa.harvard.edu/}} team for the Ariel Data Challenge 2023\footnote{\url{https://www.ariel-datachallenge.space/}}, where one of the models secured the top position among 293 competitors. Leveraging Normalizing Flows, our models predict the posterior probability distribution of atmospheric parameters under different atmospheric assumptions. 

Moreover, we introduce an alternative model that exhibits higher performance potential than the winning model, despite scoring lower in the challenge. These findings highlight the need to reevaluate the evaluation metric and prompt further exploration of more efficient and accurate approaches for exoplanet atmosphere spectra analysis.

Finally, we present recommendations to enhance the challenge and models, providing valuable insights for future applications on real observational data. These advancements pave the way for more effective and timely analysis of exoplanet atmospheric properties, advancing our understanding of these distant worlds.
\end{abstract}

\section{Introduction}

    \subsection{Retrieving Exoplanet Atmospheric compositions}
     The most commonly used method to study the atmospheric composition of an exoplanet is called \textit{transmission spectroscopy}, which consists of measuring how light from the host star gets absorbed by the planetary atmosphere during a transit, i.e. when the planet crosses in front of the disk of its star \citep{Seager_Sasselov2000}. The resulting transmission spectrum contains signatures of the chemical composition of the planetary atmosphere, its temperature, as well as clouds and hazes \cite[e.g.][]{Sing2016}. 
     
     Over the past two decades we have slowly collected spectra of a few tens of exoplanets, and that number will now more than triple with the advent of JWST\footnote{\url{https://webb.nasa.gov/}}. The next leap in the number of observed exoplanet atmospheres will come with the launch of the Ariel Space Mission\footnote{\url{https://arielmission.space/}} (currently planned for 2029), which will survey the atmospheres of about 1000 planets. 

     Retrieving the atmospheric properties of exoplanets from their transmission spectra is challenging. There have been many efforts in this direction, including \citep[e.g.][]{Line2016, Barstow2017, MacDonald2017, Tsiaras2018, Pinhas2018, Fisher2018, Brogi2019, Welbanks2019,LustigYaeger2023jwst}. Due to the high dimensionality of the parameter space and the low-resolution spectra at hand, there are degeneracies, and a deterministic answer is not usually possible or informative. Rather, a posterior probability distribution on the parameters that are consistent with observed spectra is required to understand the planet's atmosphere.  Probabilistic models to retrieve such posteriors are currently computationally demanding but until now often represented the best approach available. Consequently, a number of codes, most based on Bayesian sampling algorithms but some on machine learning too, have been developed to do this \cite[see compilation of existing codes in][]{MacDonald_2023}.  

    Bayesian sampling codes typically take a long time to run on existing computer clusters, especially when considering very complex models and high dimension parameter space, making it infeasible to use such codes to fit the spectra of hundreds of planets, over a wide range of parameters and atmospheric assumptions, in short periods of time. Our goal is to develop an efficient machine learning model for this task, that can be trained to fit spectra of planets in short time spans, therefore enabling a fast and reliable analysis of hundreds of planets spectra. Moreover, we want to develop a robust model that can retrieve the planet's composition under varying atmospheric assumptions.

    \subsection{The Ariel Data Challenge}

        \paragraph{} To enable a fast and reliable analysis of exoplanet atmospheric spectra from the upcoming Ariel Space telescope \cite{ARIEL_chemical}, the \textit{Ariel Data Challenge Organising team } have established a Data Challenge. The objectives from the 2022 Data Challenge are described in \cite{challenge_2022_def}. For the 2023 Data Challenge, the difficulty was increased by making the training set sparser, the planetary observations more difficult, and the chemistry more non-linear. This article outlines our model that was ranked first in the 2023 Data Challenge and won the challenge finals.

        \paragraph{The dataset: } 
        The generation of the 2023 Data Challenge dataset relies on the TauREx3 radiative transfer code \cite{TauREx_intro}, and follows a similar procedure that was used for the 2022 dataset \cite{challenge_2022_dataset}. The target parameter space contains 7 dimensions: Radius of the exoplanet, Temperature, and abundances of H$_2$O, CO$_2$, CO, CH$_4$ and NH$_3$. To predict the distributions of these parameters, we have access to the spectra (intensity and noise for 52 bins in the infrared domain), and to auxiliary variables of the planet and its host star (the stellar distance, mass, radius, and temperature, and the planet's mass, orbital period, distance from the star, and surface gravity). 
        
        We have two distinct sets of target parameters to train on: 1) the input parameters (i.e. the parameters that were given to the TauREx forward model to produce the spectrum), later referred to as the \textit{input parameters} or 2)  a set of samples, with weights, obtained by running a Nested Sampling (NS) with the forward model, later referred to as the \textit{set of NS samples}. We also have access to the quartiles tables (i.e. the value, for each individual parameter, of the 16$^\text{th}$, 50$^\text{th}$ and 84$^\text{th}$ percentiles of the NS samples, on each spectrum), but were not utilized in this work, since they were only a summary of the NS samples data.
        
\begin{figure}[!h] 
            \begin{center}
                \includegraphics[scale=0.4]{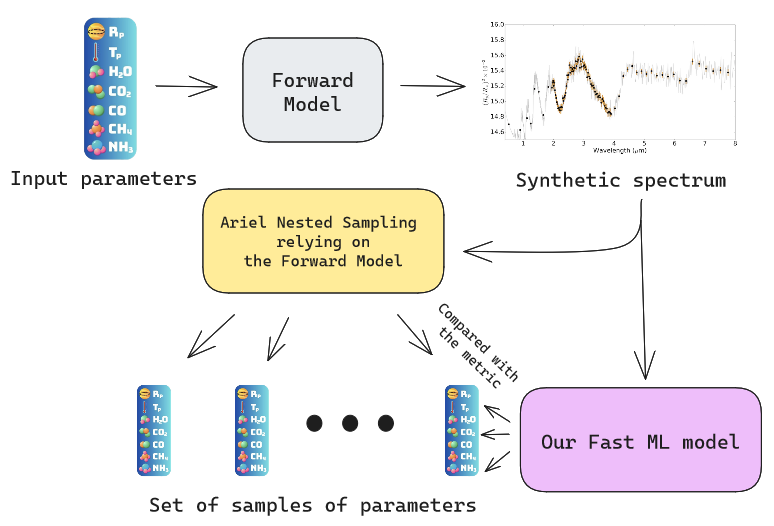}
            \end{center}
            \caption{Outline of the Ariel Challenge dataset generation.  The input parameters are used to generate the synthetic exoplanet spectra and noise arrays. A Nested Sampling is carried by the \textit{Ariel Data Challenge Organising team } on these synthetic ideal spectra, producing samples of the target parameters. Our Machine Learning model aims to reproduce the samples of parameters of the Nested Sampling. See text for more details.}
            \label{Ariel_dataset}
        \end{figure}

        The dataset generation is explained in Figure \ref{Ariel_dataset}: Some input parameters are sampled from a prior, a forward model (TauREx) computes the ideal synthetic spectrum using these, an NS algorithm with this forward model was run to obtain samples, and the task is to replace the Nested Sampling by a fast machine learning model. The total training set contains 41423 pairs of spectra/input parameters (of which, 4000 were used for testing) but only 16.3\% (i.e., 6766) NS samples were provided (of which, 1000 were used for testing). The leader-board score is determined from a test set of 685 additional pairs of spectra/parameters, while 1545 are used for the final scoring.

        \paragraph{Noise: } A noise array is provided alongside each sample exoplanet spectrum to be used for training and testing.  The exoplanet spectra are considered ideal, that is to say the  noise has not been included. This means that these spectra are not representative of expected observations, and thus our winning model \textit{as it is} will not be directly operable on real data. Here, we also present an alternative model trained on spectra with added noise (hereafter ``noised spectra''), to showcase a more realistic application of the model to real data. 

        \paragraph{Scoring metric: } The Data Challenge test set was made of unseen spectra generated with slight modifications of the forward model that were unknown to participants. These modifications included the addition of clouds, addition of another chemical species (i.e., anomalous spectral absorption), stellar activity, and addition of a non-uniform temperature profile. We submitted our predictions of samples for each test spectrum. The score is divided as follows: 80\% for a 2-sample Kolgomorov-Smirnov (K-S) test, and 20\% for a spectral score comparing the reconstructed spectrum to the original one. We focused our efforts on minimizing the K-S test. The K-S test considers the maximum difference between the cumulative distribution functions of the two samples for each target parameter. Therefore, it only measures the fitting of the marginals, not of the joint probability.
        The spectral score was measuring the distance between the original spectrum and the one reconstructed from the atmospheric parameters predicted. We decided to only focus on the K-S test because of its weight in the score and because it was easier to monitor and to improve in span of time we had for the challenge.

\section{Methods}
    Our python codes can be found in our github repository\footnote{\url{https://github.com/AstroAI-CfA/Ariel_Data_Challenge_2023_solution}}, and a summary of the important characteristics is given below.
    
    \subsection{Normalizing flows: the backbone of our model}
    
\begin{figure}[!h] 
            \begin{center}
                \includegraphics[scale=0.2]{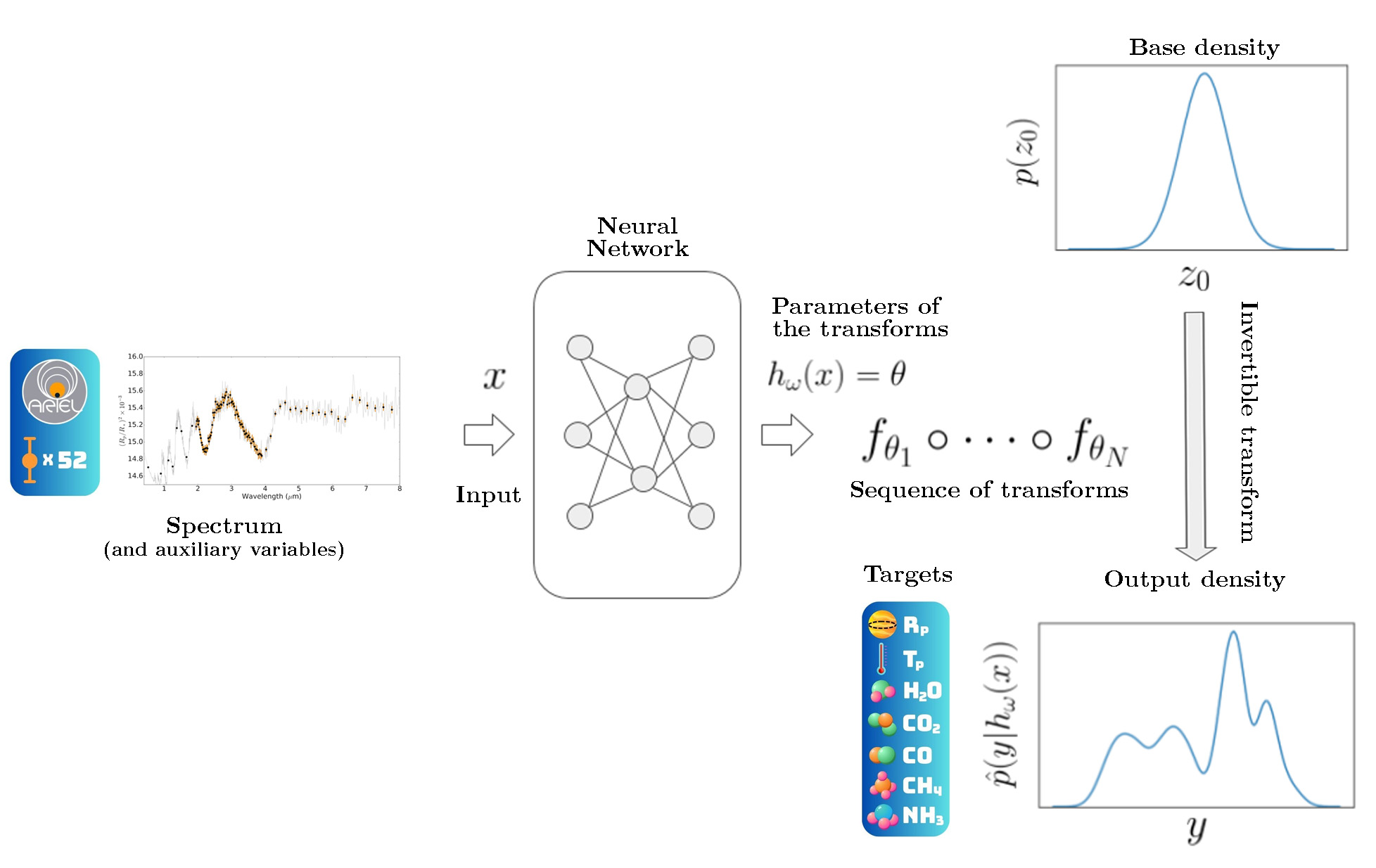}
            \end{center}
            \caption{Neural Spline Flow concept: A neural network $h_\omega$ takes a spectrum (and auxiliary variables) $x$ as input to predict the parameters $\theta$ of a sequence of invertible differentiable transforms $f$. These transforms $f_{\theta_1},...,f_{\theta_N}$ are used to turn a base density (Gaussian in our case) $p(z_0)$ into the target density $p(y|x)$. See text for more details. Figure inspired by \cite{NFconcept}}
            \label{NF_concept}
        \end{figure}

        \paragraph{Definition of Normalizing Flows: }
        Normalizing Flows have gained popularity in the field of simulation based inference, as they offer flexible and easily trainable models for posterior distributions \cite{rezende2016variational}. In particular, they have been widely applied to solve inverse problems in astrophysics \citep{Wang_2023, Vasist_2023}. Normalizing flows are defined as transformations that convert a simple probability distribution, such as a standard normal distribution, into a more complex distribution through a series of invertible and differentiable mappings \cite{Kobyzev_2021}. In our study, we utilized Neural Spline Flows to model the posterior distribution of atmospheric parameters given the observed spectra. Neural Spline Flows, introduced by Durkan et al. \cite{durkan2019neural}, employ monotonic rational-quadratic splines to model the invertible mapping, and neural networks to predict the necessary parameters of these transformations. To implement Neural Spline Flows, we utilized the Zuko python package \cite{Rozet_Zuko_Normalizing_flows_2022}. Figure \ref{NF_concept} illustrates the architecture of a Neural Spline Flow.

        \paragraph{Modelling posterior distributions with Normalizing flows: }
        The main objective in retrieving the atmospheric composition of exoplanets is to generate a posterior distribution of atmospheric parameters, which includes temperature and mass fractions of various species, based on an observed spectrum of the atmosphere. Normalizing flows can be used to approximate the probability distribution of the atmospheric parameters given the observed spectra. The model is trained using Variational Inference (VI) by minimizing the Kullback Leibler (KL) divergence \citep{10.5555/2981345.2981371} between the normalizing flow modelled posterior, $p_\phi(\theta|X)$ and the true posterior, $p(\theta|X)$, where $\theta$ are the atmospheric parameters, $X$ the observed spectra and $\phi$ the parameters of the normalizing flow that we are optimizing. The optimal parameters are found by minimizing the loss

        \begin{equation}
            \label{eq:loss}
            \phi^\star = \argmin_{\phi} \mathbb{E}_{p(X)} \left[ \mathrm{KL} (p(\theta|X) || p_\phi (\theta|X)) \right] = \argmin_{\phi} \mathbb{E}_{p(\theta, X)} \left[ - \log p_\phi(\theta|X) \right].
        \end{equation}
        
        Note that $p(\theta|X)$ is amortized, meaning that the same normalizing flow can obtain samples for any observed spectrum $X$ and does not need to be retrained.
        
        In essence, the normalizing flow maps the spectrum data to the corresponding distribution of atmospheric parameters, allowing for a comprehensive characterization of the exoplanet's atmosphere. Normalizing flows have been first used to model atmospheric retrieval in \cite{Vasist_2023}. 

        \paragraph{Independent normalizing flows: } Since 80\% of the challenge metric is the Kolgomorov-Smirnov test on the marginals cumulative distribution functions (CDF),the model is only evaluated on the posterior marginals and not on the joint. Therefore, we use independent Neural Spline Flows for each target parameter, in order to improve our performance on the challenge. That is, instead of having a single Normalizing Flow transforming an $n$-dimensional standard normal into the posterior distribution of the target parameters, we have $n$ Normalizing Flows, one for each target parameter (i.e. 7). We discuss the consequences of this choice in the Discussion \ref{metric_implications}.

    \subsection{Preprocessing the data}

\begin{figure}[!h] 
            \begin{center}
                \includegraphics[scale=0.55]{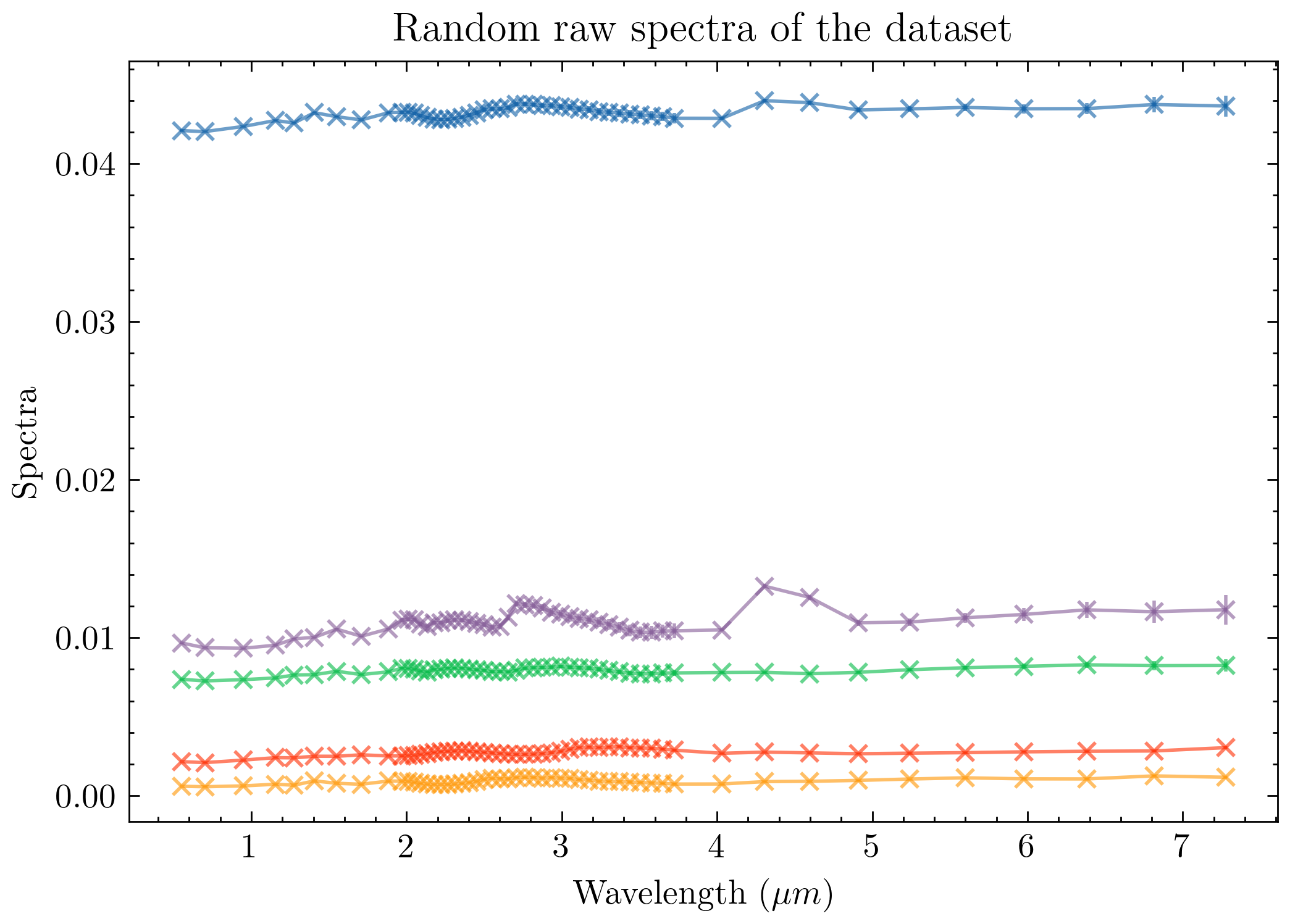}
                \includegraphics[scale=0.55]{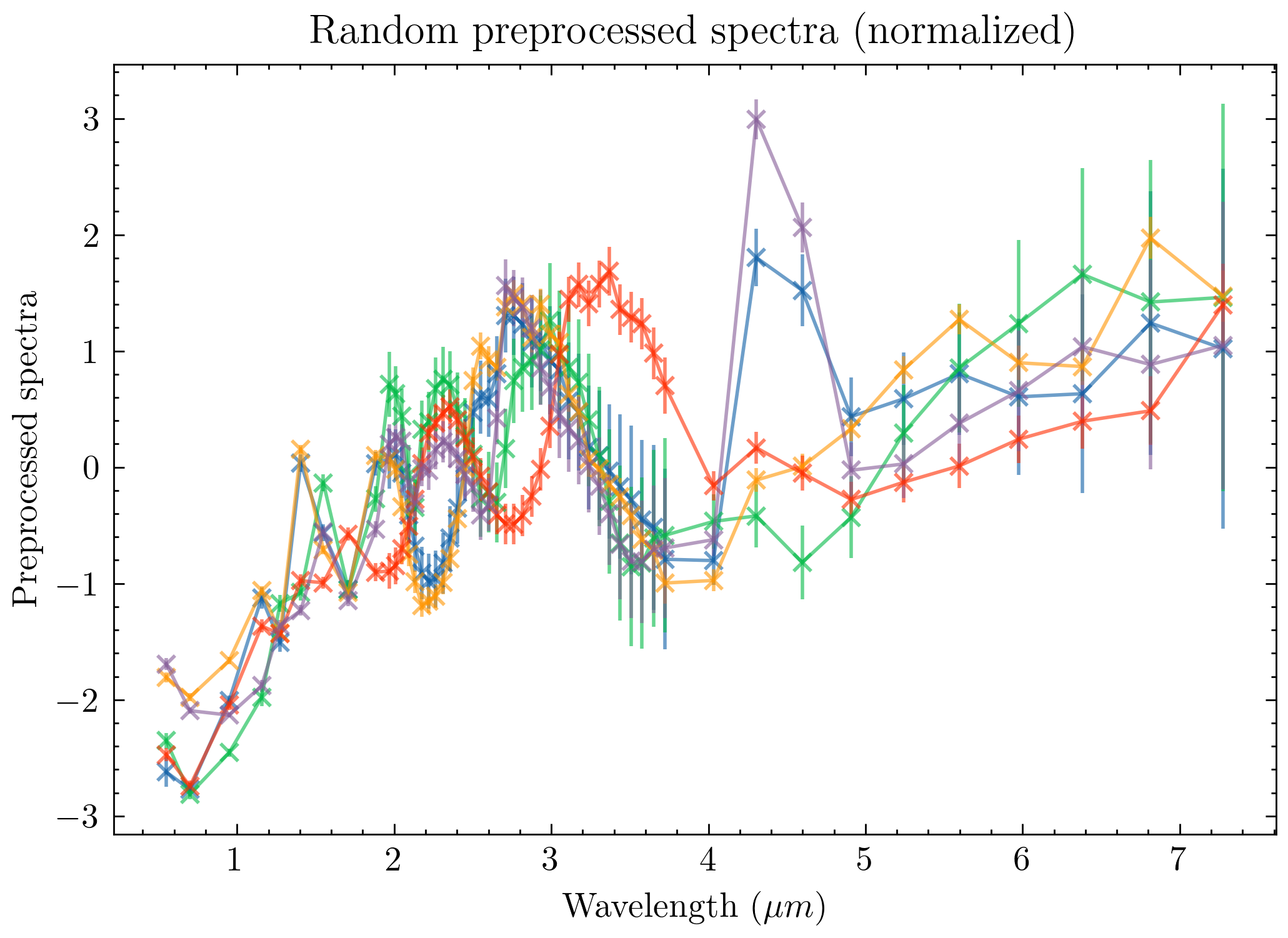}
            \end{center}
            \caption{Comparison of the some random raw spectra (left) from the ARIEL test set, and their normalized version after preprocessing (right), the colors serve as a mapping. Normalization highlights the small variations that encode the atmospheric composition.}
            \label{spectra}
        \end{figure}

        \paragraph{Transforming the spectra: }
        A spectrum can be decomposed into a constant term, mainly related to the ratio between the radius of the planet and that of the star, and variations, mainly related to the atmospheric composition. If we consider the raw spectra, their intensity seems dominated by the constant term. This is shown in Figure \ref{spectra}. We thus decided to highlight the variations by re-scaling each spectrum individually: we set their mean to zero and their standard deviation to one. We save the mean and standard deviation as new features for our model. Therefore the model can more easily extract the information concerning the atmospheric composition contained in the shape of the spectrum (its variations).

        \paragraph{Transforming the auxiliary variables and target parameters: }
        We normalized the targets parameters, the auxiliary variables and the noises too. We normalized every feature by subtracting the minimum on the training dataset and dividing by the difference between the maximum and minimum. 
        We also used two estimators of the Radius of the planet: one derived from the mean of the spectrum, and the other one from the surface gravity and planet mass. We used a combination of both depending on the data availability to create a combined estimator. 
        
        The models just had to predict the offset between the real radius and the combined estimator. Therefore, the estimator was used as a bootstrap for predicting the radius. We also provided the combined estimator, the two estimators, and a Boolean representing the data availability to our models as additional input features.

    \subsection{Training the models and Hyperparameters optimization}

        \paragraph{Loss: logarithmic probability or distance to the cumulative distribution function?}
        Even though the metric of the challenge is the 2-sample K-S test, we decided to train our models with the logarithmic probability as the loss, since Normalising Flows can predict the logarithmic probability of any set of targets. The 2-sample K-S test would have taken too long to compute and backpropagate because of the number of samples per spectrum (around $2000$). Instead, we use the negative weighted logarithmic probability (i.e. $-\sum_i\alpha_i\log(p_\phi(\theta_i|X))$ where $\alpha$ and $\theta$ are the weights and targets of the samples given by nested sampling, $X$ is the spectrum analysed and $p_\phi$ the prediction of the Normalising Flow). Note that, as stated in Eq.~\eqref{eq:loss}, this loss is equivalent to the forward Kullback–Leibler divergence on the samples (see also equation (13) of \cite{papamakarios2021normalizing}). Alternatively, when using independent Normalizing Flows, one could use the 1-sample K-S test as the loss since we can derive an analytic formula of the predicted CDF using the composition of the transformations of the flow and the $erf$ function (from the base Gaussian density).
        
        To highlight the limitations of the challenge's metric and to showcase an application of the model that is closer to real data, we also present two alternative models, both with Normalizing Flows that model the joined posterior distribution and that are trained on the input parameters instead of using the NS samples, that is the loss is the logarithmic probability on the input parameters used to generated the spectra in the first place. See \ref{model_tested} for their exact definition.
        
        \paragraph{Hyperparameters optimization: }
        For our winning and alternative models, we ran hyperparameters optimization with Optuna \cite{optuna_2019}. In the case of the winning model, since we have seven independent Normalizing Flows, we ran seven independent hyperparameters optimisations, one for each target parameter.
        
        The hyperparameters concerned were: the number of transformations of the flows (ranging from 1 to 20), the number of bins of the splines (from 3 to 40), and the number of hidden layers (from 1 to 5) and features of the contextual neural network (from 1 to 500). For the winning model, there are 7 independent ensembles of the 10 best Normalising Flows, giving a total of 70 sets of hyperparameters. We could not identify any specific pattern in these sets, the hyperparameters are distributed throughout the range of optimisation, with slight differences depending on the target parameter of the independent flow.
        
        \paragraph{Ensembling the best models: }
         Once the hyperparameters optimization was complete, we ensemble the $10$ best models to reduce model's errors and increase robustness. That is, we use the $10$ best models to predict the samples for the submission, with a number of samples allocated to each model depending on its performance (the better the model, the more samples it has). For the winning model, we used as weight for the number of samples allocated the following expression: $\frac{1-KS_i}{\sum_i(1-KS_i)}$ where $KS_i$ is the K-S test of the model $i$ on the local test set (i.e. the fraction of the available training set that we used for testing).
         
         In the following, we will use winning model and alternative models to refer to the winning independent ensembles of Normalising Flows and the alternative ensembles of joined Normalising Flows.

\section{Results}

    \paragraph{The models tested: }
    \label{model_tested}
    We compare the results of three models: A model optimized to win the challenge, referred as the winning model; a model optimizing the logarithmic probability on the input parameters of the forward model, as opposed to the NS samples, and trained with the original ideal spectra, referred to as the alternative ideal model; and a model optimizing the logarithmic probability on the input parameters and trained with noised spectra, referred as the alternative noised model.
    
    The winning model is a set of independent ensembles of Normalizing Flows, one ensemble per target parameter, which hyperparameters optimization objective was the K-S test. Even though the Normalizing Flows were still trained with the logarithmic probability, the metric maximized by the hyperoptimization was the K-S test, computed on the local test set.
    
    Both alternative models are ensembles of joined Normalizing Flows, 7-dimensional flows. Their hyperparameters optimization objective was the logarithmic probability of the input parameters, computed on our local test set. 
    
    The metrics of our three models are summarised in the table of results (Figure \ref{Table_of_results}).

    \subsection{Local performance}

        \paragraph{Logarithmic probability: } The mean logarithmic probability gives, to within a constant, the forward Kullback–Leibler divergence. We can use the mean logarithmic probability on the input parameters as a metric to compare models on their ability to reproduce the posterior of atmospheric parameters given the spectrum. The logarithmic probability on ideal spectra is computed with ideal spectra as inputs, and measures the capability of retrieving the parameters when provided the ideal synthetic spectra. Instead, the logarithmic probability on noised spectra is computed with noised spectra as inputs, with noise sampled from the provided uncertainties levels. The logarithmic probability on noised spectra is much closer to what we would expect from a real usage of the models, as it measures the ability of retrieving the input parameters from a realistic simulated observation.
        
        As can be seen in Figure \ref{Table_of_results}, the best model for the logarithmic probability on ideal spectra is the alternative ideal model, as expected, but it has a huge drop when considering noised spectra. While the alternative noised model is, although less precise on ideal spectra, more stable under the addition of noise. This demonstrates that the alternative noised model is learning to model the uncertainty, contrary to the alternative ideal model.

        \paragraph{Kolgomorov-Smirnov test: } We compute the local K-S test with the set of NS samples as the target distribution. We used the Scipy implementation \texttt{scipy.stats.ks\_2samp} to compute the K-S test between the predicted samples and the set of NS samples. This K-S test measures the maximum difference between the cumulative distribution functions of the marginals. Once again, the K-S test on ideal spectra is computed with ideal spectra as inputs for the models, while the K-S test on noised spectra uses noised spectra as inputs. The K-S test on ideal spectra composes 80\% of the challenge score, with the difference that the challenge score is measured on a slightly different test set.\\
        Both winning model and alternative noised model have similar K-S test on noised spectra, even if the winning model has the best K-S test on ideal spectra. 

        \paragraph{Coverage of the posterior distributions: } To evaluate the accuracy of the model's posterior uncertainty estimates, we can look at the coverage of the posterior distributions, shown in Fig.~\ref{Coverage_plots}. We construct it by defining a range of percentiles, spanning from 0 to 1, each corresponding to a specific confidence interval from our estimated posterior distribution. These serve as our nominal coverage levels. We then empirically assess the proportion of instances where the true parameter value falls within these confidence intervals, forming our empirical coverage levels. In a perfectly calibrated model, the proportion of times the true parameter value falls within a given confidence interval (say $95\%$) is exactly equal to the nominal level ($95\%$ in this case), illustrating that our uncertainty estimates are accurately calibrated. Points below this line would indicate undercoverage, meaning the true parameter falls within the confidence interval less frequently than expected. Conversely, points above the line would indicate overcoverage, where the true value falls within the confidence interval more frequently than anticipated.

        In Fig.~\ref{Coverage_plots}, we see that the NS and winning model posteriors are both showing deviations from a perfect coverage, suggesting that there may be some bias or calibration issue with the NS. It could be coming from the difference of prior used in the generation of the input parameters and in the setup of the NS. Nonetheless, the winning model reproduces the same behavior, since it was trained to mimic the NS. On the other hand, the alternative noised model produces posteriors that seem to be better calibrated with respect to the input parameters.

        \begin{figure}[!h] 
            \begin{center}
                \begin{tabular}{l c c c}
                	Metrics  & \textbf{Winning } & \textbf{Alternative } & \textbf{Alternative } \\
                             &  \textbf{model} & \textbf{ideal} & \textbf{noised} \\
                	\hline\\
                	Logprob (ideal spectra)      & $3.16$ &  $5.02$ & $2.86$ \\
                	\hline\\
                    Logprob (noised spectra)     & $1.03$ & $-6.30$ & $2.48$ \\
                    \hline\\
                    K-S test  (ideal spectra)     & $0.11$ &  $0.59$ & $0.42$ \\
                    \hline\\
                    K-S test  (noised spectra)    & $0.45$ &  $0.70$ & $0.47$ \\
                    \hline\\
                    Challenge score              & $688.13$ &  $457.83$ & $577.32$ \\
                \end{tabular}
            \end{center}
            \caption{Comparison of our two main models on the metrics considered. Logarithmic probability has to be maximized, while K-S test has to be minimized.}
            \label{Table_of_results}
        \end{figure}

    \subsection{Scores in the Ariel Data Challenge}

        \paragraph{Effects of the various improvements: } We started with a model made of Independent Normalizing Flows, so one flow per target parameter (7 in total), with fixed hyperparameters. It scored 679 on the challenge leaderboard. Then we ran an hyperparameters optimization with the objective our local K-S metric, took the ten best models for each parameter and grouped them in an ensemble. It scored 688, showing that running an hyperparameters optimization with ensembling can result in a $1.33$\% performance increase. In the meantime, our alternative ideal model, an ensemble of joined Normalizing Flows trained on the input parameters (and still ideal spectra), scored 458, showing that the use of NS samples as targets and independent Normalizing Flows produced $48.3$\% score increase.

    \subsection{Comparing the winning model to alternatives}

\begin{figure}[!h] 
            \begin{center}
                \includegraphics[scale=0.6]{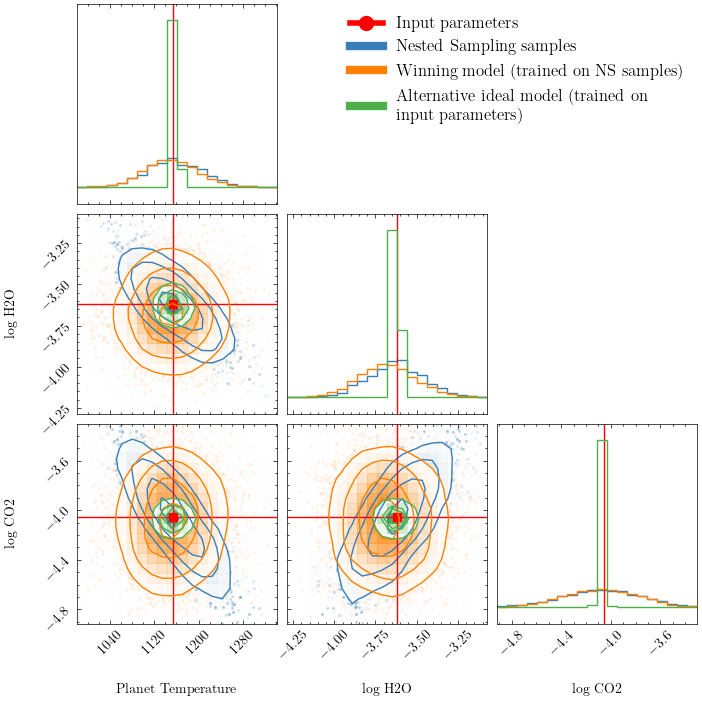}
            \end{center}
            \caption{Comparison of the posteriors of T, H$_{2}$O and CO$_{2}$ for a previously unseen random ideal spectrum. Examples of full corner plots can be found in the appendix of \cite{ASTRO_paper}}
            \label{Posterior_comparison_ideal}
        \end{figure}

\begin{figure}[!h] 
            \begin{center}
                \includegraphics[scale=0.6]{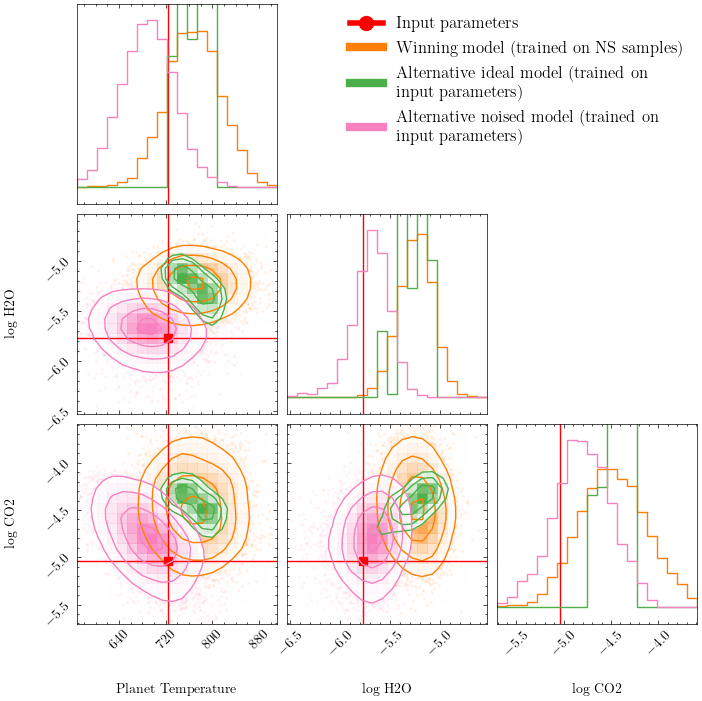}
            \end{center}
            \caption{Comparison of the posteriors of T, H$_{2}$O and CO$_{2}$ for a previously unseen random noised spectrum (different from Fig.~\ref{Posterior_comparison_ideal}). We did not include the NS posteriors, since they were retrieved from ideal spectra (not noised spectra). Examples of full corner plots can be found in the appendix of \cite{ASTRO_paper}}
            \label{Posterior_comparison_noised}
        \end{figure}

        \paragraph{Distribution of parameters: } To get a better understanding of the behaviors of the models, we can compare the posteriors obtained for some random spectra never seen by the models. We can do the comparison with ideal spectra as inputs, to reproduce the setup of the challenge, or with noised spectra, to see how the models would behave on realistic simulated spectra.
        An extract of a posterior comparison on ideal spectra is shown in Figure \ref{Posterior_comparison_ideal}, and on noised spectra in Figure \ref{Posterior_comparison_noised}.
        The posterior of the alternative ideal model in Figure \ref{Posterior_comparison_ideal} shows that training a model with the input parameters without adding noise to the spectra leads to unrealistic overconfident predictions.
        To emphasize on the inability of the alternative ideal model to deal with observations, we compare its predictions to the input parameters with both ideal or noised spectra in Figure \ref{Heatmap}.

\section{Discussion}
%TODO: make it clearer how important it is to test on noised spectra, clarify the explanations on noise

    \subsection{Implications of the Ariel Data Challenge dataset and metric}
    \label{metric_implications}
    The metric and the dataset have decisive impacts on the models endorsed by the challenge. Challengers will strive to maximize their scores using any means possible, sometimes at the expense of achieving the true objective. In our case, we opted to employ independent Normalizing Flows to improve our results, even though it led to a loss of correlation between parameters.
    
\begin{figure}[!h] 
        \begin{center}
            \includegraphics[scale=0.23]{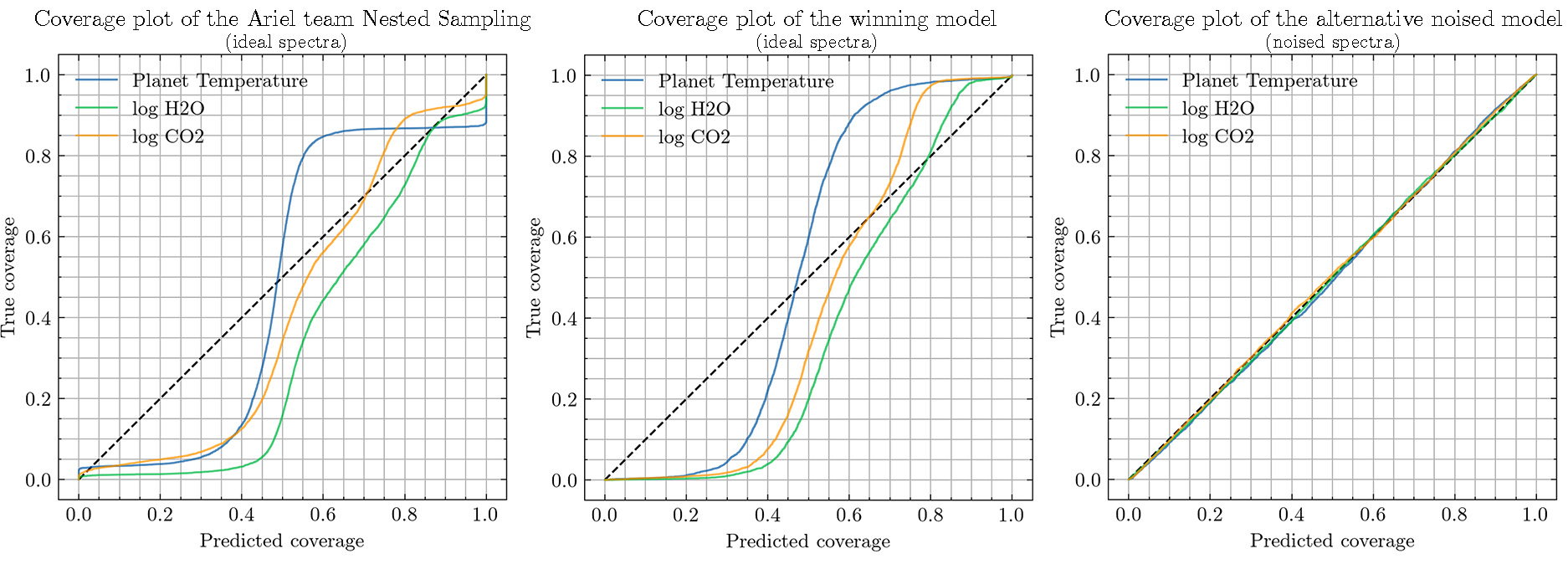}
        \end{center}
        \caption{Coverage plots of the Nested Sampling on ideal spectra (left), our winning model on ideal spectra (center), and the alternative noised model on noised spectra (right). We are plotting the fraction of predicted samples inferior to the input parameter (x-axis) as a function of the fraction of the input parameter from the minimum to the maximum (y-axis). It is equivalent to the composition of the inverse estimated CDF and the true CDF. The deviations around the straight line indicate the bias or misestimation of the variance.}
        \label{Coverage_plots}
    \end{figure}

    The main metric of the challenge, the Kolgomorov-Smirnov test accounting for 80\% of the score, is measuring the maximum distance of the CDF of the marginals, between the proposed samples and the set of samples obtained by running an NS. We identified three major limitations of this approach:
    \begin{enumerate}
        \item \textbf{Joint probabilities:} Since the test is on the marginals only, it is dismissing all information concerning the joint probabilities. Models on this metric are not compelled to learn any relationship between parameters. Such relationships could result from degeneracies between parameters. We can see in Figure \ref{Posterior_comparison_ideal} that the NS posterior has a diagonal term between the temperature and the H$_2$O mass fraction that the winning model does not learn. Losing that information does not cost anything with the K-S test.
        \item \textbf{Noise treatment:} The models were tested on ideal spectra only. They are thus not tested on their ability to analyse real spectra. As we can see in Figure \ref{Posterior_comparison_noised} and in the table of results \ref{Table_of_results}, the alternative noised model perform better on noised spectra. We do not expect our winning model to be capable of retrieving any coherent atmospheric parameters from real observed spectra.
        \item \textbf{Biased dataset:} Using the set of NS samples instead of the input parameters as targets may introduce bias in the model, or at least some approximation. As we can see in the coverage plot of the set of NS samples (Figure \ref{Coverage_plots}), the set of NS samples seem not well calibrated. In fact, the model will not learn to retrieve the true posteriors, but to mimic the behavior of the Nested Sampling, and thus any bias or approximation would be reproduced. We think that this step is unnecessary since the model could learn directly from the input parameters.
    \end{enumerate}

    We can see that the model that won the challenge is reproducing the exact same behaviors as the Nested Sampling. This is shown in the comparison of distributions on ideal spectra Figure \ref{Posterior_comparison_ideal}, where the winning model marginals are fitting the NS samples marginals; and in the coverage plots Figure \ref{Coverage_plots}, where the coverage produced by the NS and winning model have the same shape. Yet, we can also see that our alternative ideal model, even though having a lower challenge score, seems to be giving more precise and exact posteriors on ideal spectra. This is directly caused by the treatment of the noise: no noise is directly added to the spectrum (the spectrum are the ideal results of the input parameters), but an array of synthetic noise is provided in addition to the spectrum. The NS is set to consider this noise as real noise (it considers that the real spectrum is in the range given by the noise), so it outputs large uncertainties on the target parameters; while our alternative ideal model is only trained to retrieve the input parameters, which it can do very well since the spectra are not noised. 
    
\begin{figure}[!h] 
        \begin{center}
        \includegraphics[scale=0.30]{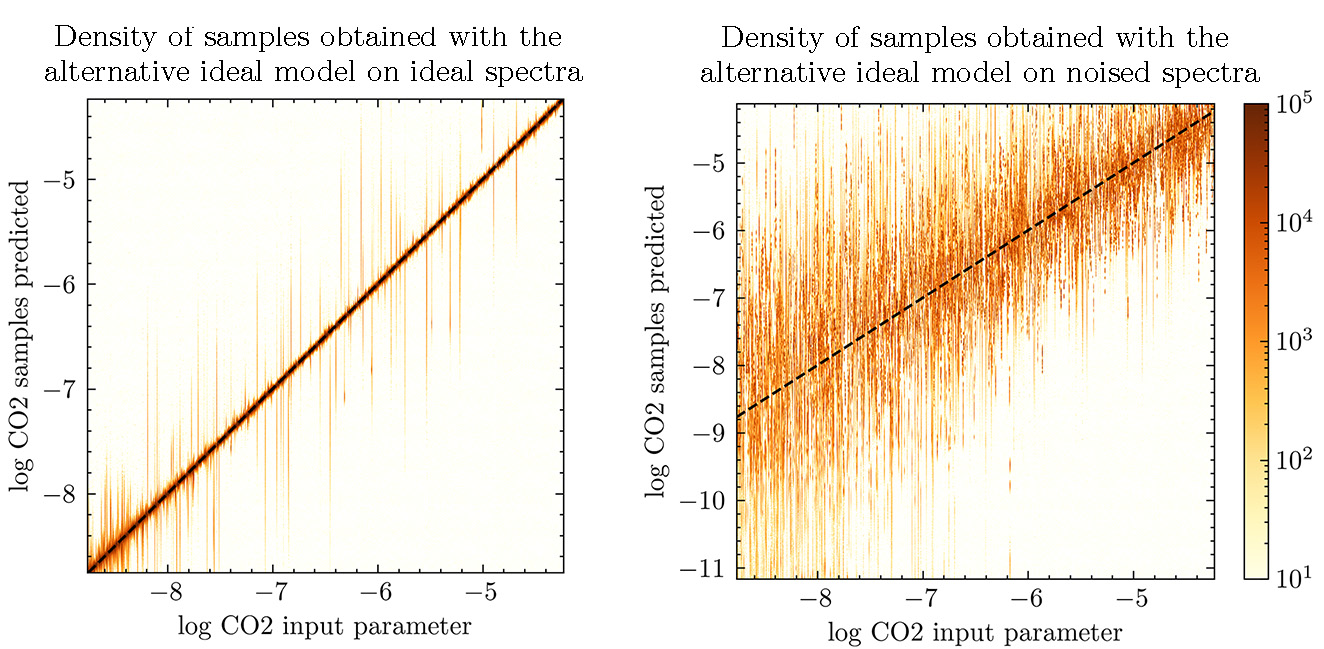}
        \end{center}
        \caption{Samples of CO$_{2}$ compared to the input CO$_{2}$ for alternative ideal model on ideal spectra (left) and noised spectra (right). The black dotted line is the identity (the target). The complete heatmaps for all parameters of the other models can be found in the appendix of \cite{ASTRO_paper}.}
        \label{Heatmap}
    \end{figure}

    The alternative ideal model is not considering the noise array as a real noise. We can see how adding noise to the spectra deteriorates the performance of the alternative ideal model on Figure \ref{Heatmap} and Figure \ref{Posterior_comparison_noised}. That is why we also present an alternative noised model capable of dealing correctly with uncertainties since it is trained on noised spectra and not ideal spectra. Figure \ref{Posterior_comparison_noised} demonstrates that it performs significantly better than our other models on noised spectra.

    \subsection{Proposing an alternative metric and test set} 
    We acknowledge that designing a metric for a data challenge is a highly complex task. In this section we propose potential improvements to the metrics based on our results. 
    
    The metric has to account for accuracy, the calibration of the posteriors, in one single number while being light in computational demand. Also, one has to ensure that the models can be used to produce samples. In this respect, the K-S test enabled dealing with samples, ensuring that the model would be able to provide samples, and measure the accuracy and calibration of the marginals compared to the set of NS samples. Yet, it has the flaws highlighted in \ref{metric_implications}.
    
     Lueckmann et al. \cite{lueckmann2021benchmarking} proposed a set of benchmarks for simulation-based inference models. It appeared to us that using the logarithmic probability on the input parameters may account for the accuracy and calibration of the complete posterior while not introducing any bias. This is the metric we used for our alternative model. But for it to make the model exploit the uncertainties on the spectra (and propagate them correctly to the parameters), the uncertainties need to be meaningful uncertainties, that is uncertainties giving the probability distribution of the ideal spectra: an ideal spectrum should follow a Gaussian probability distribution of mean the noised spectrum and of standard deviation the uncertainty.

     We do not recommend to provide a training set of noised spectra (the teams can add noise as their will), but we insist on the fact that scoring the models on noised spectra is mandatory if the goal pursued is to promote models capable of dealing correctly with uncertainties.

    \subsection{Application of our model to real data}
    The scores on the metrics, and the behavior of the posterior comparison plots, demonstrate the capability of Normalizing Flows to reliably and expeditiously retrieve from the infrared spectrum the posterior distributions of exoplanets atmospheric parameters for synthetic ideal data. Yet some issues should be addressed before being able to deploy our models:
    \begin{enumerate}
        \item \textbf{Observational uncertainties:} The models are not able  to handle observational uncertainties. They need to be trained on noised data with a noise model incorporating all the phenomenon at play in exoplanet atmosphere spectroscopy.
        \item \textbf{Variable binning:} The models only accept data with a fixed binning along the wavelength, making it viable for only one data pipeline.
        \item \textbf{Various forward models:} There exist several forward models, each of them able to generate different spectra for the same input parameters, depending on the atmospheric assumptions considered.
    \end{enumerate}
    For the first issue, the $\text{SBI}^{++}$ model \citep{Wang_2023} implements an elegant solution, allowing it to cope with out-of-distribution noise, that is noise levels not seen in the training set, thus making the model more reliable when used on real data.
    
    The second point could be resolved by changing the architecture of the contextual neural network, with a model accepting any binning; or with a common latent space and modules for each input encoding into this latent space; or could be ignored if considering a standardized pipeline, such as the raw data directly.
    
    Regarding different atmospheric assumptions, a model trained on varying assumptions would be able to marginalize over them. This problem is also related to model misspecification. Even if we can model all atmospheric assumptions, there could still be missing pieces in our understanding of exoplanet atmospheres that would lead to a biased inference of the planet's composition. It is therefore important to work on ways to summarize the observations in a robust manner, by understanding how the simulated spectra differ from the observed one. See \citep{huang2023learning} for some recent work on robust summarization.

\section{Conclusion}
 
With the availability of advanced telescopes like JWST and the upcoming Ariel space telescope, obtaining spectra of exoplanet atmospheres has become feasible, presenting an opportunity to unlock valuable information about their composition and structure.

In this research article, we addressed the challenging task of retrieving the atmospheric composition of exoplanets from their observed spectra using machine learning techniques. We applied our model to the Ariel Data Challenge, where we ranked first among 293 teams. The dataset utilized the TauREx3 radiative transfer code and focused on a 7-dimensional target parameter space, including exoplanet radius, temperature, and abundances of key compounds like H$_2$O, CO$_2$, CO, CH$_4$, and NH$_3$. Our goal was to replace laborious and time-consuming Bayesian sampling algorithms with a faster and more efficient machine learning model, that can easily generalize to different atmospheric assumptions.

To achieve our objective, we utilized Normalizing Flows, a versatile and easily trainable model to describe complex probability distributions. Specifically, we employed Neural Spline Flows, which leverage rational-quadratic splines and a neural network to describe the parameters for the transformations. This approach enabled us to map the spectrum data to the corresponding probability distribution in the target parameter space, encompassing temperature and mass fractions of various species.

While the challenge metric was based on the Kolgomorov-Smirnov test and spectral score, we found certain limitations in this approach. It only measured fitting of the marginals, ignoring joint probabilities and probably introducing bias from the use of Nested Sampling samples as targets. Nevertheless, our winning model effectively mimicked the NS behavior, indicating its capability to produce samples and approximate the posteriors.

We proposed an alternative metric, the logarithmic probability on the input parameters of the forward model, which better accounted for accuracy and calibration without introducing bias. However, testing the models on noised spectra is essential to validate their ability to comprehend uncertainties and handle real data effectively.

Our models demonstrated strong potential for retrieving atmospheric parameters from synthetic ideal data, but further work is required to address observational uncertainties, variable binning, and various forward models. Solutions such as $\text{SBI}^{++}$ model, latent space models, and regularization terms were suggested to enhance robustness and adaptability.

In conclusion, our research showcases the promising capabilities of Normalizing Flows in retrieving exoplanet atmospheric composition, paving the way for future advancements in the field of astrophysics and exoplanet research.\\\\

\textit{Software: }
    PyTorch \cite{paszke2019pytorch},
    Zuko \cite{Rozet_Zuko_Normalizing_flows_2022},
    Matplotlib \cite{Hunter:2007},
    Corner \cite{corner},
    Scipy \cite{2020SciPy-NMeth},
    Numpy \cite{harris2020array},
    Pandas \cite{mckinney-proc-scipy-2010},
    SciencePlots \cite{SciencePlots},
    Excalidraw \cite{excalidraw}

\section*{Acknowledgments}
%What should we put here ???
This team was put together, led and supervised by AstroAI at the Center for Astrophysics | Harvard \& Smithsonian. Mayeul Aubin and Cecilia Garraffo were partially supported by the Director's Office at the Center for Astrophysics | Harvard \& Smithsonian. AstroAI thanks the Harvard Data Science Initiative for their support. Jeremy J. Drake was supported by NASA contract NAS8-03060 to the {\it Chandra} X-ray Center during the course of this research. Iouli Gordon, Robert Hargreaves, and Vladimir Makhnev were supported by NASA PDART grant 80NSSC20K1059 throughout this work. \\
We thank the NSF AI Institute for Artificial Intelligence and Fundamental Interactions (IAIFI) for providing computational resources through the Faculty of Arts and Science Research Cluster (FASRC) of Harvard.

\clearpage 
\bibliography{refs} % Entries are in the refs.bib file

\end{document}